\documentclass[%
 reprint,
 superscriptaddress,
 amsmath,amssymb,
 aps,
]{revtex4-2}

\usepackage{graphicx}% Include figure files
\usepackage{float}
\usepackage{dcolumn}% Align table columns on decimal point
\usepackage{bm}% bold math
\usepackage[%  
    colorlinks=true,
    pdfborder={0 0 0},
    citecolor=blue,
    linkcolor=blue
]{hyperref}
\begin{document}

% \preprint{APS/123-QED}

\title{Large Language Model-Assisted Superconducting Qubit Experiments}

\author{Shiheng Li}
\affiliation{Department of Physics, University of Chicago, Chicago, IL 60637, USA}
\author{Jacob M. Miller}
\affiliation{Department of Physics, University of Chicago, Chicago, IL 60637, USA}
\author{Phoebe J. Lee}
\affiliation{Department of Physics, University of Chicago, Chicago, IL 60637, USA}
\author{Gustav Andersson}
\affiliation{Pritzker School of Molecular Engineering, University of Chicago, Chicago, IL 60637, USA}
\author{Christopher R. Conner}
\affiliation{Pritzker School of Molecular Engineering, University of Chicago, Chicago, IL 60637, USA}
\author{Yash J. Joshi}
\affiliation{Pritzker School of Molecular Engineering, University of Chicago, Chicago, IL 60637, USA}
\author{Bayan Karimi}
\affiliation{Pritzker School of Molecular Engineering, University of Chicago, Chicago, IL 60637, USA}
\affiliation{Pico group, QTF Centre of Excellence, Department of Applied Physics, Aalto University School of Science, P.O. Box 13500, 00076 Aalto, Finland}
\author{Amber M. King}
\affiliation{Pritzker School of Molecular Engineering, University of Chicago, Chicago, IL 60637, USA}
\author{Howard L. Malc}
\affiliation{Pritzker School of Molecular Engineering, University of Chicago, Chicago, IL 60637, USA}
\author{Harsh Mishra}
\affiliation{Pritzker School of Molecular Engineering, University of Chicago, Chicago, IL 60637, USA}
\author{Hong Qiao}
\affiliation{Pritzker School of Molecular Engineering, University of Chicago, Chicago, IL 60637, USA}
\author{Minseok Ryu}
\affiliation{Pritzker School of Molecular Engineering, University of Chicago, Chicago, IL 60637, USA}
\author{Xuntao Wu}
\affiliation{Pritzker School of Molecular Engineering, University of Chicago, Chicago, IL 60637, USA}
\author{Siyuan Xing}
\affiliation{Department of Mathematics, University of Chicago, Chicago, IL 60637, USA}
\author{Haoxiong Yan}
\altaffiliation[Present address: ]{Applied Materials, Inc, Santa Clara, CA 95051, USA}
\affiliation{Pritzker School of Molecular Engineering, University of Chicago, Chicago, IL 60637, USA}
\author{Jian Shi}
\affiliation{Department of Materials Science and Engineering, Rensselaer Polytechnic Institute, Troy, NY 12180, USA}
\author{Andrew N. Cleland}
\email[Corresponding author: ]{anc@uchicago.edu}
\affiliation{Pritzker School of Molecular Engineering, University of Chicago, Chicago, IL 60637, USA}

\date{\today}

\begin{abstract}
Superconducting circuits have demonstrated significant potential in quantum information processing and quantum sensing. Implementing novel control and measurement sequences for superconducting qubits is often a complex and time-consuming process, requiring extensive expertise in both the underlying physics and the specific hardware and software. In this work, we introduce a framework that leverages a large language model (LLM) to automate qubit control and measurement. Specifically, our framework conducts experiments by generating and invoking schema-less tools on demand via a knowledge base on instrumental usage and experimental procedures. We showcase this framework with two experiments: an autonomous resonator characterization and a direct reproduction of a quantum non-demolition (QND) characterization of a superconducting qubit from literature. This framework enables rapid deployment of standard control-and-measurement protocols and facilitates implementation of novel experimental procedures, offering a more flexible and user-friendly paradigm for controlling complex quantum hardware.
\end{abstract}
 

\maketitle

%\tableofcontents

\section{Introduction}

The field of superconducting quantum circuits is evolving rapidly from proof-of-concept demonstrations to a leading platform for large-scale quantum information processing \cite{google2025quantum,ibm2025qdc}. Beyond the primary goal of universal quantum computation, superconducting qubits have also catalyzed breakthroughs in various research frontiers of hybrid systems \cite{clerk2020hybrid} by coupling with acoustics \cite{qiao2023splitting,qiao2025acoustic}, spins \cite{zhu2011coherent}, semiconductors \cite{scarlino2019coherent}, optics \cite{mirhosseini2020superconducting}, and so on \cite{wu2024modular,wu2025mitigating}. As circuit designs continue to diversify and hardware capabilities improve at an accelerating pace, efficiently controlling, calibrating, and measuring these quantum systems becomes increasingly laborious and challenging.

Artificial intelligence (AI) systems based on large language models (LLMs) have advanced rapidly over the past few years \cite{minaee2024large}. Significant effort has been invested in using the ability of LLMs to reason, code, and write in scientific research. Innovative designs have been proposed by LLMs in biology and medicine \cite{swanson2025virtual}, while autonomous experiments have been demonstrated in chemistry, quantum, and material science laboratories \cite{tom2024self,cao2025automating}. Inspired by these successes, we extend LLM capabilities into superconducting qubit experiments. However, challenges exist for the direct application of LLMs in the laboratory. Unlike everyday software environments where methods like the model context protocol (MCP) \cite{hou2025model} excel, scientific research thrives on the dynamic evolution of hardware and experimental procedures. Furthermore, floating-point data such as microwave scattering parameters are often confusing for LLMs, and experimental practice often requires human interpretation of inconclusive results.

If these challenges can be overcome, superconducting qubit experiments can then be performed by LLMs, as their purely digital operation removes the requirement for mechanical interventions. Superconducting circuit systems, governed by the principles of circuit quantum electrodynamics (cQED), utilize Josephson junctions to achieve the anharmonicity required for qubit operation. While the foundational physics of these devices is detailed extensively in the literature \cite{blais2021circuit,krantz2019quantum}, their operational framework is particularly accessible for AI systems. Once a device is cryogenically cooled, all control and measurement can be mediated exclusively through electronic signals, which are typically managed via high-level programming languages such as Python, falling directly into the expertise of LLMs. If an LLM is allowed to interact with the lab infrastructure directly through Python, we can exploit the abilities of LLMs to pursue quantum experiments.

In this work, we build a framework that applies LLMs to superconducting qubit experiments. We first introduce our lab setup and highlight several open-source Python packages. We then establish the workflow of our AI system. We abandon the conventional tool-based agent architecture and implement a succinct structure to develop schema-less new tools on the fly. Supported by a knowledge base, the AI system can rapidly deploy standard control-and-measurement protocols, while allowing customization and innovation in experimental methods using natural language. Finally, we demonstrate the AI system and provide functional examples, one where the AI automates resonator characterization and a second where it conducts a quantum non-demolition (QND) characterization experiment. The latter is implemented by the LLM using only a general-purpose laboratory knowledge base combined with a description in a published journal article \cite{hazra2025benchmarking}.

\section{Methods}

\subsection{Lab Setup}

\begin{figure*}
\includegraphics[width=\textwidth]{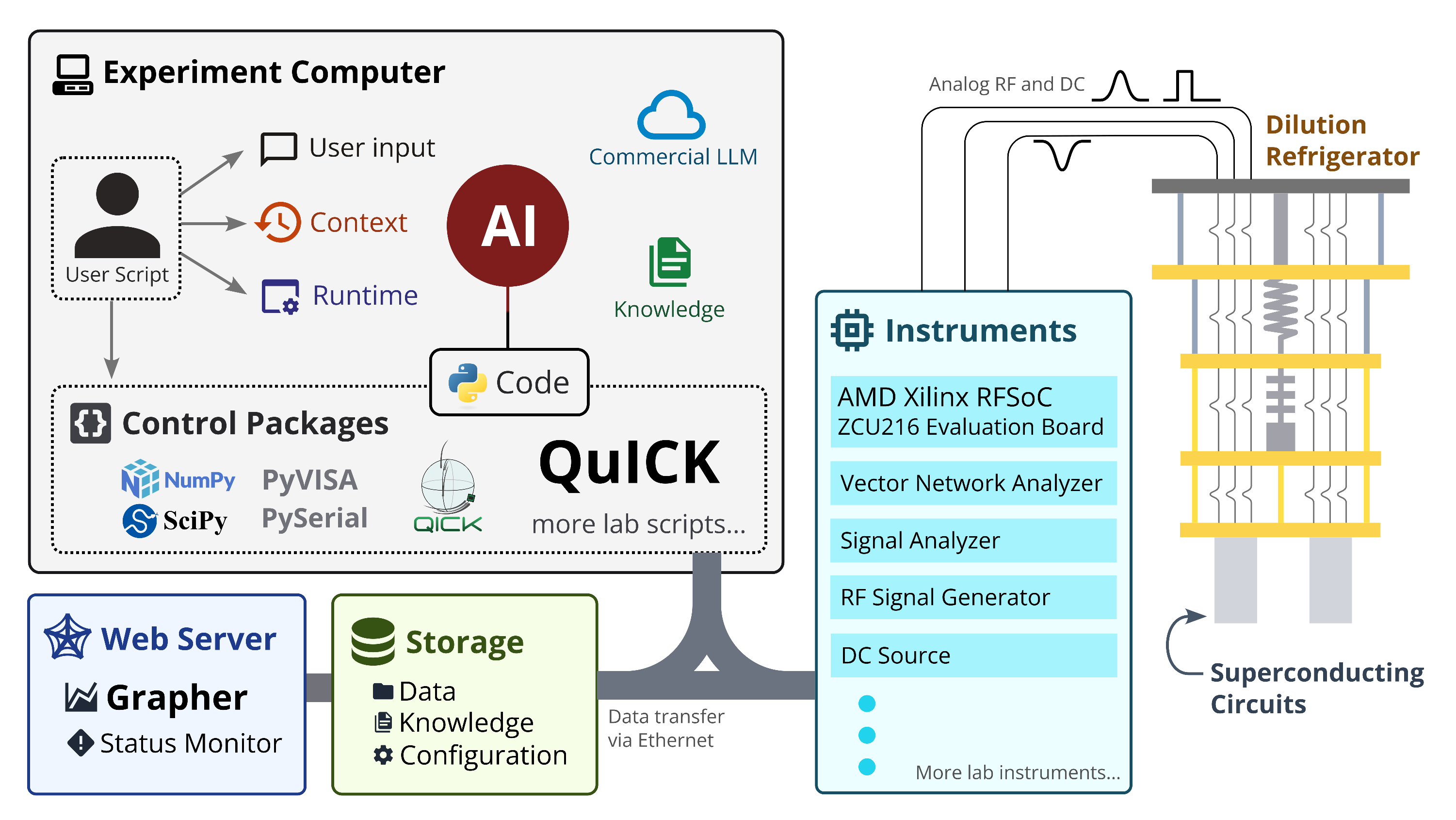}
\caption{\label{fig:lab}Schematic overview of the experimental setup. The measurement system is organized into a multi-layer architecture that bridges high-level user intent with physical quantum circuits. Via the experiment computer, both users and AI systems can interact with specialized control packages, most notably the open-source package QuICK \cite{github_quick}. The experiment computer is connected to a local area network (LAN), through which commands are transmitted to a suite of instruments, which synthesize and detect analog RF and DC signals routed to a dilution refrigerator (DR) to interact with superconducting circuits at cryogenic temperatures. Experimental progress and results are managed by a centralized storage server and visualized in real-time through the open-source Grapher software \cite{github_grapher}. In practice, multiple sets of computers, instruments, and DRs are connected to the same LAN to allow convenient and simultaneous experiments across the laboratory.}
\end{figure*}

Our experimental setup is schematically illustrated in Fig.~\ref{fig:lab}. The architecture is designed in a multi-layer format to incorporate a variety of lab instruments and research experiments, while maintaining both convenience and flexibility in daily practice. On the physical connection layer, experiment computers and electronic instruments are connected to a local area network (LAN) via Ethernet connections. A centralized storage server is also connected to this LAN and mounted on all experiment computers as a network file system, and a dedicated web server is deployed with our open-source Grapher software \cite{github_grapher}, providing real-time data visualization and monitoring. During an experiment, commands from the experiment computer are transmitted to a set of electronic instruments (Xilinx RFSoC evaluation boards \cite{stefanazzi2022qick}, vector network analyzers (VNAs), voltage sources, etc.) that generate analog radio frequency (RF) and direct current (DC) signals. The signals are routed to a dilution refrigerator (DR) where they interact with a superconducting circuit operating at cryogenic temperatures ($\sim 10$ mK). After the interaction, RF signals are routed back to electronic instruments and sampled by an analog-to-digital converter (ADC), from which digital data are transferred to the experiment computer via the LAN. The computer can then process the data and save it in the storage server to be viewed by the user using the Grapher software \cite{github_grapher}. Details regarding the superconducting circuits \cite{blais2021circuit,krantz2019quantum}, analog signal processing, and cryogenic technology \cite{krinner2019engineering} are beyond the scope of this work. In practice, multiple sets of experiment computers and lab instruments are connected to the same LAN, allowing plug-and-go usage of the lab equipment. This setup makes it possible for a research group to conduct multiple experiments simultaneously, providing a straightforward path for an AI to play a role in experiments without making dramatic changes to the lab infrastructure.
 
The multi-layer architecture extends to the software side on the experiment computer. A user can use Python scripts to interact with the AI system, described below. Both the user and the AI system can then perform an experiment using Python code, with an abstract software layer that includes a series of control packages to communicate with the instruments and handle the data processing. 

In addition to existing Python packages such as PyVISA \cite{grecco2023pyvisa}, we develop an open-source package, QuICK \cite{github_quick}, which is a wrap of the Quantum Instrumentation Control Kit (QICK) package \cite{stefanazzi2022qick} for the control of the Xilinx RFSoC evaluation boards (QICK boards). QuICK is designed to provide a simple interface in the quantum control code. It optimizes installation and configuration, while providing a wide range of useful functions. Most critically, QuICK introduces a declarative syntax to specify a pulse sequence for qubit control and measurement, and immediately plots previews of the pulse sequence. Using flat and simple syntax, both human users and AI systems can easily produce customized control sequences. Combined with powerful helper functions, QuICK covers all the basic control-and-measurement requirements using the Xilinx RFSoC evaluation board. On top of pulse-level control, QuICK also includes commonly-used experimental routines and a series of automatic single-qubit tune-up scripts. Independent of the AI system, QuICK serves as a convenient control package for superconducting qubit experiments.

\subsection{AI System}

\begin{figure*}
\includegraphics[width=\textwidth]{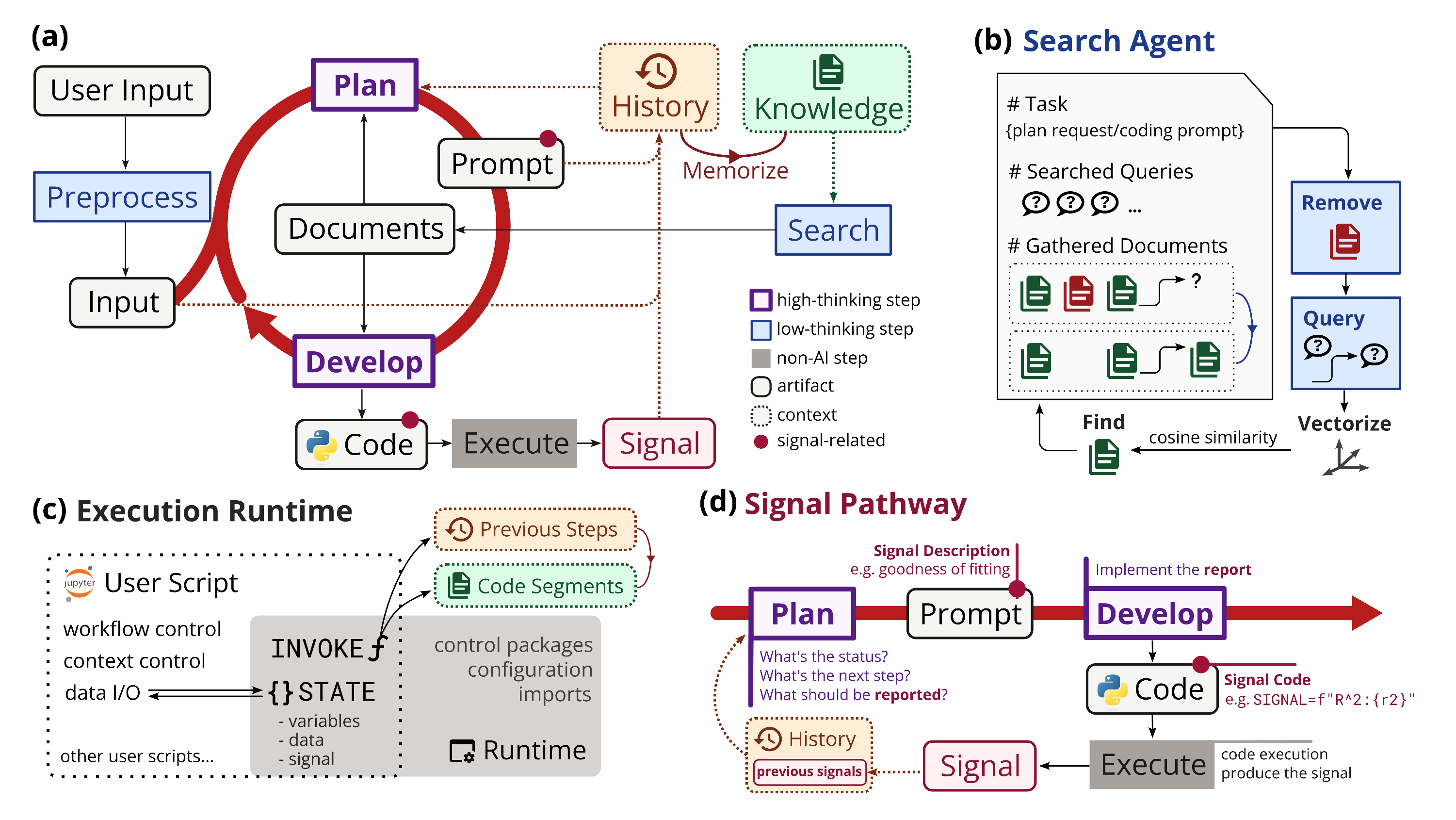}
\caption{\label{fig:hal}Architecture of Heuristic Autonomous Lab (HAL). (a) Workflow: The system operates in a cycle with multiple steps, supported by long-term (green) and short-term (amber) contexts. High-thinking steps (purple) involve complex tasks including planning and development, while low-thinking steps (blue) handle routine tasks like searching. (b) Search Agent: Iterative retrieval-augmented generation (RAG) process to accurately find relevant documents from the knowledge base using vectorization and cosine similarity, effectively handling references among documents by iteration. (c) Execution Runtime: A sandbox environment, where the generated code is invoked, maintains a persistent state across executions and allows the invocation of other codes. (d) Signal Pathway: A feedback mechanism to close the information loop, allowing the AI system to acquire critical information about the execution results.}
\end{figure*}

\paragraph{Workflow Overview} On top of the control packages, we introduce our AI system, the Heuristic Autonomous Lab (HAL), driven by Gemini, a leading commercial LLM \cite{google2026gemini}. Figure \ref{fig:hal} illustrates the HAL system. Here we use Gemini 3 Flash (\texttt{gemini-3-flash-preview}) as the primary LLM, but we believe this method is generalizable to other LLMs. As shown in Fig.~\ref{fig:hal}(a), the general workflow is a cycle with two major steps: \textit{plan} and \textit{develop}. Instead of a tool-based agent architecture, we repetitively ask the AI system two key questions: what to do next, and how to do it. The \textit{plan} component (\textit{planner}) addresses the first question, given a full history of the previous steps and the relevant knowledge documents. It produces a detailed textual prompt, analogous to that of a human manager. This prompt is fed into the \textit{develop} component (\textit{developer}), which acts like a professional programmer and implements Python code from the prompt, addressing the second question. This Python code is handed to HAL Runtime, a dedicated component that executes the code in a controlled environment and manages the execution input and output, the results of which are reported as a signal (discussed below). The \textit{planner} and \textit{developer} are powered by LLM calls with high-thinking configuration \cite{google2026gemini}.

User inputs and the knowledge base are mapped into the system via low-thinking steps. Raw user input is processed by a \textit{preprocess} step which refines the user narrative and, more importantly, separates queries from commands. Queries are requests expressed as a question and expecting a literal response. When the LLM sees a query, it has a tendency to provide an answer instead of performing an action, jeopardizing the main HAL workflow, which is designed for commands and actions. For example, ``how to restart the computer'' will result in a text answer, while ``restart the computer'' will perform an action. Thus, we redirect the queries to a separate \textit{answer} component, which is a context-supported chatbot not shown in the figure. In contrast, commands are merged into the history and often initiate the cycle. In parallel, relevant knowledge is acquired from the database by a \textit{search} component, which we discuss later.

Context is the most crucial piece in the HAL system, providing integration of the different components, where we distinguish short-term from long-term context. Short-term context is the step history, including the previous inputs, prompts, and signals. This history is provided to the \textit{planner} to help it understand the status of the current session and make a decision about what to do next. Long-term context is the knowledge stored in the system. The knowledge base is a set of text documents permanently stored in the storage server, which includes tutorials, application programming interface (API) documents, standard protocols, code examples, etc. To support semantic search, an embedding vector is computed by the Gemini embedding model for each document. A subtle but important feature is memorization, which is a human-directed self-improvement mechanism. When a task is successfully completed, the working prompt and code can be reviewed by the user and permanently converted into code examples in the knowledge base for future reference. This conversion from short-term to long-term context enables HAL to improve itself via ``practice''. In short, we ``teach'' the system by providing knowledge documents, as if we are teaching a student who never forgets.

\paragraph{Search Agent} To effectively provide accurate long-term context for the other components, we have developed an iterative search agent for the knowledge base. The traditional retrieval-augmented generation (RAG) method \cite{gao2023retrieval} often fails to exclude irrelevant documents or incorporate Cross-referencing among relevant documents. The lateral problem is essential to the heuristic knowledge base, as human-written documents often refer to other documents. These referred documents may be semantically irrelevant to the search query and thus missing from the search results. We attempted to solve this issue using a multi-turn tool-calling agent, but this suffers from attention loss caused by long inputs \cite{laban2025llms}. We instead implemented the iterative RAG agent shown in Fig.~\ref{fig:hal}(b): The agent maintains a structured state comprising the original task, a list of previously searched queries, and a collection of gathered documents. In each iteration, the LLM is instructed to complete two steps. First, it identifies irrelevant documents. Second, it analyzes the task and the relevant documents to generate new queries for any missing information. These queries are vectorized by the Gemini embedding model to find new candidate documents using cosine similarity within the vector space of the knowledge base. The newly retrieved documents are added to the gathered set, and the iteration repeats until no new queries are generated, or until the maximum iteration number ($\sim$ 5) is reached. The two steps of identifying irrelevant documents and raising queries are sufficiently simple for LLMs operating in the low-thinking configuration, resulting in high-quality search results with low latency and cost. This \textit{search} component is called by the \textit{plan}, \textit{develop} and \textit{answer} components with slightly different instructions focusing on high-level plans, API documents, and general knowledge, respectively. The gathered documents are then attached to the instructions for other components.

\paragraph{Execution Runtime} The execution runtime environment (Fig. \ref{fig:hal}(c)) makes the generated code executable, reusable, and flexible for both human users and the AI system. Upon user configuration, HAL can automatically execute the generated Python code with all the control packages imported, or present the code to a human user for manual execution. The two execution environments share two extra global variables, \texttt{STATE} and \verb|INVOKE|. The \textit{developer} is given all the runtime information so that the resulting code can use the global imports and variables and can be executed by both HAL and the user. By saving the generated code, a user can reuse the code in future experiments with flexible customization. Moreover, the re-usability and flexibility are enhanced by the two global variables: \texttt{STATE} and \texttt{INVOKE}. \texttt{STATE} is a Python dictionary that can be modified by both the generated code and the user scripts, establishing a blackboard method to share runtime data. For example, users can place experimental parameters in \texttt{STATE}, and HAL can read these parameters, conduct experiments, and write the measured data in \texttt{STATE}. This shared blackboard makes it possible to reuse the generated code as a function with flexible inputs and outputs. \texttt{INVOKE} is a Python function that initiates the execution of other Python codes, involving code from previous steps and examples from the knowledge base. When necessary, the \textit{planner} can prompt the \textit{developer} to re-run previous steps, potentially with updates to \texttt{STATE}. In this case, the \textit{developer} can generate code with \texttt{INVOKE(step)} without reproducing the code used in the previous steps. We also grant the \texttt{INVOKE} function direct access to the code available in the knowledge base. With these constructs, we prepare the HAL system for complicated tasks.

\paragraph{Signal Pathway} What makes the HAL system stand out from conventional tool-calling systems is the signal pathway (Fig.~\ref{fig:hal}(d)), which relieves the AI system of pre-defined response schema. Integrating execution feedback is essential to move beyond a consultative role. However, we simultaneously aim to maintain architectural flexibility by avoiding the rigid output schemas typical of conventional agentic tools. To address this issue, we design a signaling mechanism to close the information loop. Instead of a fixed response structure, the signal can comprise arbitrary text whose content is informed by the \textit{planner}, implemented by the \textit{developer}, and finally assigned in code execution. When the \textit{planner} generates the prompt for the next step, it also specifies the signal description (e.g. ``number of found resonators''), which is the expected report from the step execution. The \textit{developer} then implements the prompt and the signal description into Python code. In our implementation, the signal is always assigned to a special variable in \texttt{STATE}. After code execution, this variable is read, saving the signal value (e.g. ``Found 4 resonators'') in the history. Therefore, when the \textit{planner} refers to the short-term context, it will see the signal description in the prompt together with the signal values, effectively reporting the execution outcome of the previous steps. The signal mechanism helps the AI system avoid handling incomprehensible floating-point data and dynamically adjust actions based on experimental results. The goal of the HAL system is to let the LLM make decisions and build the required tools on demand, where the signals serve as the output of the tools and are also freely determined by the system itself.

\section{Results and Discussion}

We developed the Grapher software \cite{github_grapher}, the QuICK package \cite{github_quick}, and the HAL system implemented as a Python package \cite{github_hal}. They are open-sourced on GitHub with documentation provided.
 
With a knowledge base comprising instrumental API documents and some experimental plans, the HAL system serves as a very convenient laboratory tool. The possible tasks managed by HAL increase when we provide more documents describing a larger variety of experiments, instruments, and even maintenance tasks. With added code examples, HAL also performs tasks with diminishing errors. It is worth noting that the LLM itself (Gemini 3 Flash Preview with a knowledge cutoff of January 2025 \cite{google2026gemini}) has no knowledge of our critical control packages (QuICK package and other closed-source scripts). We also exclude API documents about human-written experimental routines from HAL's knowledge base, so HAL is forced to develop experimental scripts from the pulse sequence level. Instead, standard measurement protocols are included as experimental plan documents written in natural language. 

We next describe the abilities of the HAL system using two examples, in which HAL autonomously conducts experiments with non-trivial code implementation.

\subsection{Autonomous Experiment:\\ Resonator Characterization}

\begin{figure}
\includegraphics[width=\columnwidth]{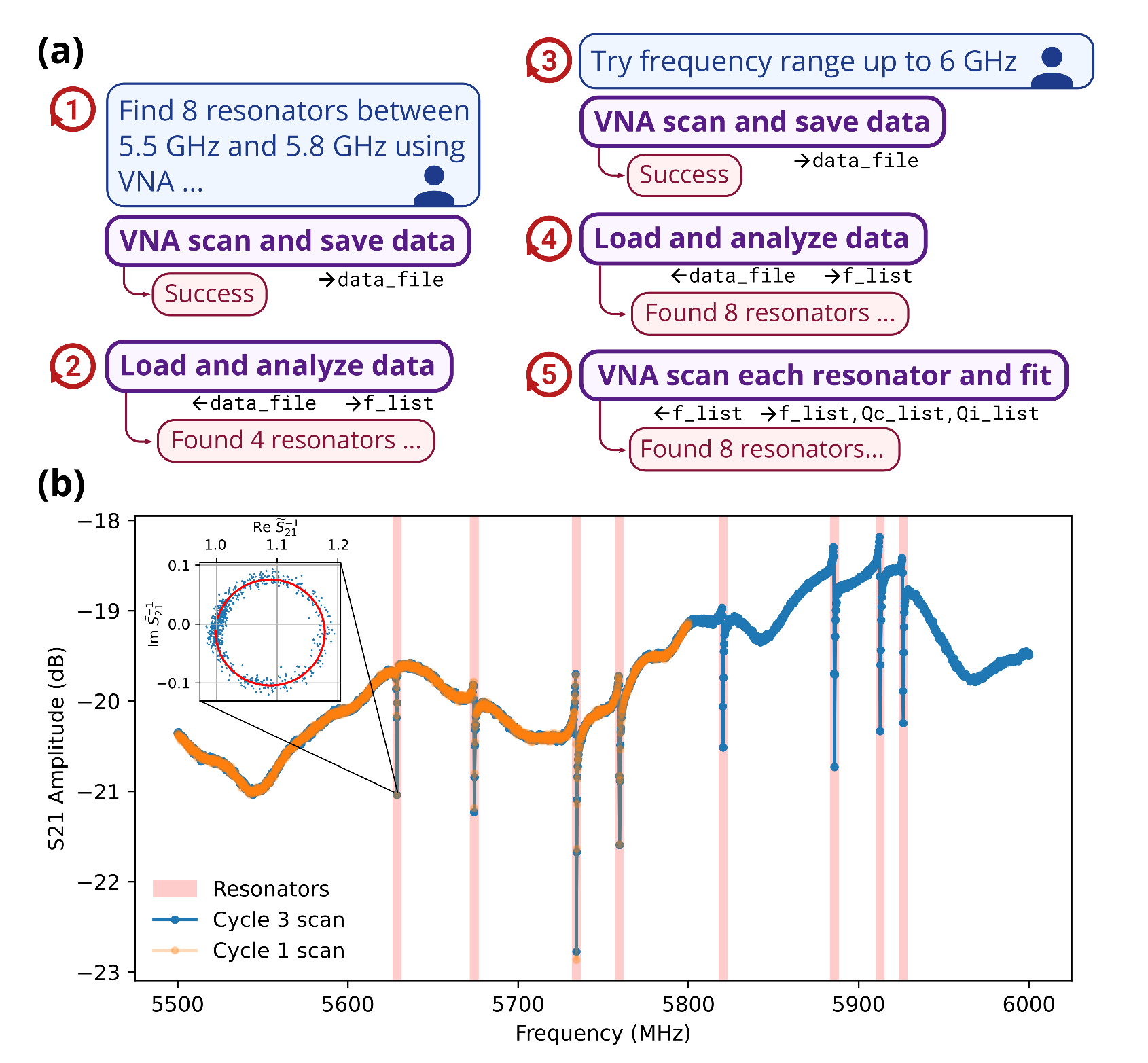}
\caption{\label{fig:resonator}Autonomous resonator characterization. (a) Summary of the user input (blue), performed actions (purple), \texttt{STATE} read and write (black), and resulting signals (red) across five cycles. Cycles 2, 4, and 5 run without any user input. (b) Measured spectrum corresponding to the vector network analyzer (VNA) scan in cycle 1 (orange) and cycle 3 (blue). Pink spans are the resonators identified after cycle 5. Inset shows a representative circle fit performed during cycle 5 \cite{megrant2012planar}.}
\end{figure}

We first showcase an autonomous resonator characterization (Fig.~\ref{fig:resonator}) as a demonstration of a standard measurement protocol with flexible customization. The standard procedure is described in a plan document written in natural language, which is placed in the knowledge base with other necessary instrumental documents. The plan document instructs the AI system to take three steps: implement a spectrum scan with a vector network analyzer (VNA), load and analyze the acquired data, and implement fine scans around the identified resonance frequencies, then fit the data to extract the coupling quality factors ($Q_c$) and internal quality factors ($Q_i$). Along with the high-level instructions, the plan document also refers to API documents on instrumental control and helper functions such as curve-fitting. Fig.~\ref{fig:resonator}(a) summarizes the sequence of HAL cycles performed in this example. It includes 5 cycles:

\begin{enumerate}
    \item A human user requests HAL to find 8 resonators, specifying instrument parameters and an intentionally narrow frequency range covering only 4 resonators. Guided by the documents retrieved from the knowledge base, the \textit{planner} generates an acquisition prompt, which the \textit{developer} follows to implement the code. During execution, the resulting spectrum (Fig.~\ref{fig:resonator}(b), orange) is saved to the storage server, its path (\texttt{data\_file}) is written in \texttt{STATE}, and a “Success” signal is returned.
    \item Without additional input, the \textit{planner} directs the \textit{developer} to analyze the stored data against known resonator features, following the plan document. The \textit{planner} also specifies that it expects a signal describing the number of identified resonances. The \textit{developer} generates code to load data using the path in \texttt{STATE} and identifies the resonances, writes the resulting frequencies (\texttt{f\_list}) to \texttt{STATE}, and signals that 4 resonances were found.
    \item At this point, we introduce human-in-the-loop \cite{mosqueira2023human} intervention. By instructing the HAL system to extend the frequency range, the \textit{planner} successfully understands the situation and prompts the \textit{developer} to take data again with an updated frequency range. Similar code implementation and data acquisition happen as in cycle 1. The acquired data is displayed in Fig.~\ref{fig:resonator}(b) in blue.
    \item The same actions are performed as in cycle 2. The resulting signal now includes 8 resonances as expected.
    \item The \textit{planner} is satisfied by the previous cycle and moves on according to the plan document. It instructs the \textit{developer} to loop through the identified resonators and perform fine scans to extract quality factors. The generated code uses the frequency list (\texttt{f\_list}) from \texttt{STATE} and acquires spectral data near each frequency. The data are fitted (inset in Fig.~\ref{fig:resonator}(b)) against the resonator model \cite{megrant2012planar} using appropriate helper functions in the QuICK package. The fitted resonator frequencies (\texttt{f\_list}) (pink spans in Fig.~\ref{fig:resonator}(b)) and quality factors (\texttt{Qc\_list}, \texttt{Qi\_list}) are saved to \texttt{STATE}.
\end{enumerate}

The relevant documents, a full transcript, and a screen record of this example can be found in the supplementary materials \cite{Supplement}. About $10^5$ input tokens and $10^4$ output tokens (including thinking tokens) are used in the process. With prepared prompts, the process takes about five minutes, including the VNA scanning time.

This example demonstrates effective assistance from HAL in superconducting circuit experiments, balancing convenience and flexibility. It indicates that the HAL system can follow a heuristic plan while allowing customization, such as back-and-forth trial of parameters. HAL is interacting with lab instruments only via generated Python code, which can be examined by humans. In addition to the human-in-the-loop control \cite{mosqueira2023human}, we also highlight the separation of data acquisition and analysis instructed by the plan document. Separating data acquisition from data analysis helps the AI system avoid inventing results because the \textit{developer} has no information about the expected analyzing steps when acquiring data. If the acquired data is hallucinated by the LLM, it is more likely to cause failure than imitating success during the data analysis step, resulting in the human user rejecting the results. Decoupling acquisition from analysis also gives the \textit{planner} a chance to catch any runtime errors in data acquisition and improve the robustness.
 
\subsection{From Journal Article to Experiment:\\QND Characterization}

\begin{figure}
\includegraphics[width=\columnwidth]{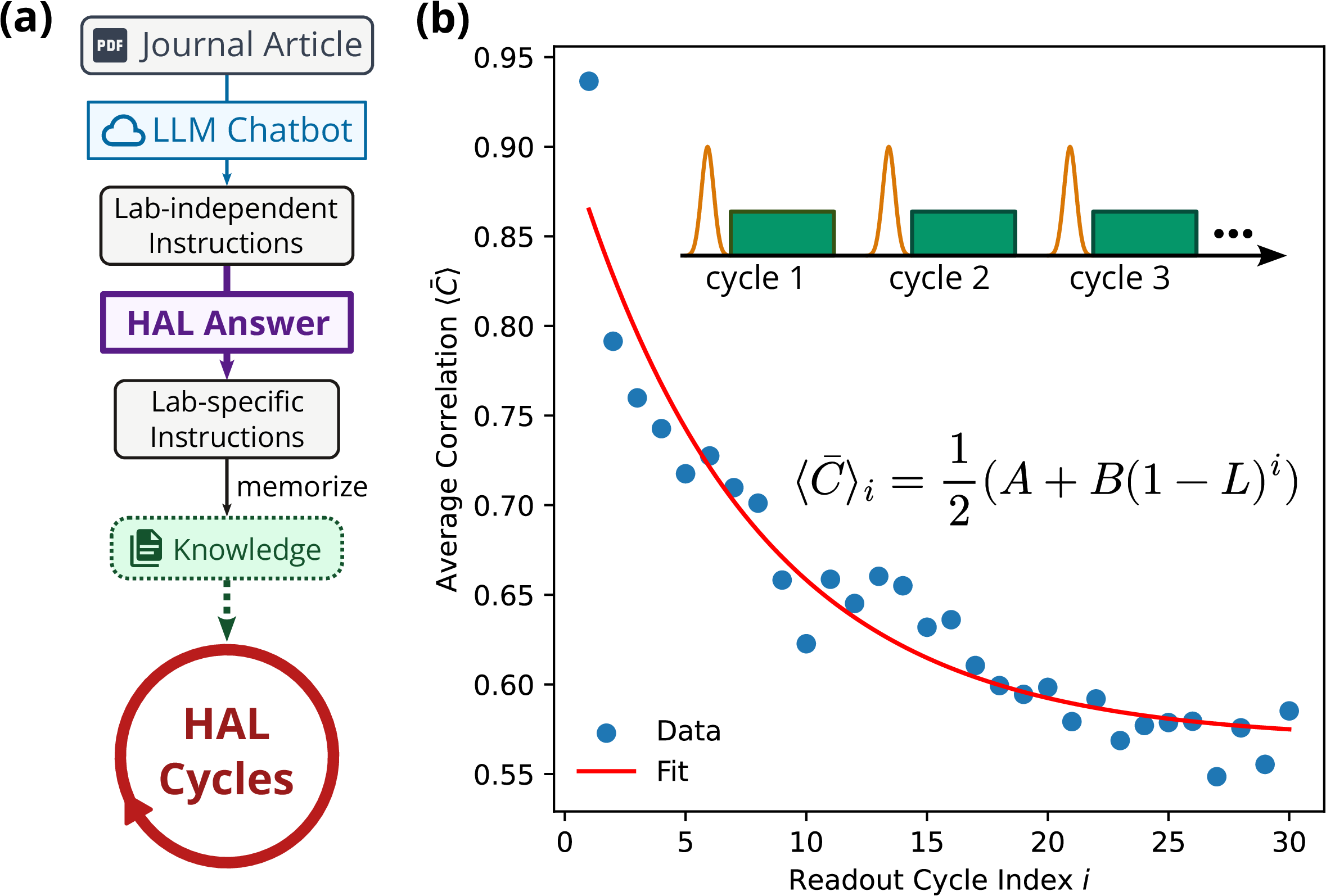}
\caption{\label{fig:qnd}Implementation of a QND characterization from a published journal article \cite{hazra2025benchmarking}. (a) Workflow for translating scientific literature into a working experiment. A journal article is processed by an LLM chatbot to generate lab-independent instructions, which are then refined into lab-specific instructions by the HAL \textit{answer} component, guiding HAL on how to conduct the experiment. (b) Experimental results of a QND characterization sequence. The main plot shows the average correlation as a function of the readout cycle index. Data (blue) is fitted against the decay model \cite{hazra2025benchmarking}, where a leakage rate from the computational basis of the qubit $L = 0.124\pm0.017$ is extracted from the fit (red). Inset illustrates the pulse sequence (not to scale), showing the randomized qubit control $\pi$ pulses (amber) and readout pulses (green).}
\end{figure}

The HAL system is not limited to human-written knowledge documents. In the second example, we demonstrate a direct reproduction of a quantum non-demolition (QND) characterization from a published journal article \cite{hazra2025benchmarking} by generating a knowledge document before starting the HAL process sequence.

The measurement described in the journal article provides a quantitative characterization of the qubit readout process that evaluates the preservation of the measured qubit state within the computational basis following qubit measurement. Measurement leakage is an undesired process in which the readout process drives the qubit outside its computational basis, making the qubit unusable in operations that follow. The readout-induced leakage benchmarking (RILB) method described in the Ref.~\cite{hazra2025benchmarking} has the pulse sequence shown inset to Fig.~\ref{fig:qnd}(b), where readout pulses (green) are interleaved randomly with qubit $\pi$ control pulses (amber). The method quantifies leakage by measuring the effectiveness of the $\pi$ pulses. In the absence of leakage, the alternation (non-alternation) of two consecutive readout results should be correlated to the existence (absence) of the $\pi$ control pulse in between the readout processes, up to Pauli errors and qubit state discrimination errors. However, leakage from a readout process can drive the qubit out of its computational basis, thereby disabling the functionality of the $\pi$ pulses, effectively destroying the subsequent correlations. This causes an exponential decay in the averaged correlation $\langle \bar{C}\rangle$ with respect to the readout cycle index $j$. With sufficient repetitions of differently randomized sequences and fitting the averaged correlation $\langle \bar{C}\rangle$ against the model (Eq.~\ref{eq:qnd}) \cite{hazra2025benchmarking}, we can extract the leakage rate $L$ of our qubit readout process:

\begin{equation}\label{eq:qnd}
    \langle \bar{C}\rangle_j = \frac{1}{2}(A+B(1-L)^j),
\end{equation}
where $\langle \bar{C}\rangle_j$ is the averaged correlation between the qubit alternating readout state and qubit $\pi$ pulse for readout cycle $j$, $A$ and $B$ are parameters representing the effect of state discrimination, state initialization and Pauli errors, and $L$ is the average leakage probability per readout (leakage rate).

We note that none of the knowledge about the QND characterization and the corresponding leakage characterization has been previously presented to the HAL system. To prepare the HAL system for this experiment, we have to generate a piece of knowledge combining the experimental method described in the article and the instrumental platform in the lab. Fig.~\ref{fig:qnd}(a) shows the workflow of knowledge preparation. It includes three steps:

\begin{enumerate}
    \item Prompt an online LLM chatbot (Gemini Thinking \cite{google2026gemini_web}) to extract the experimental procedure from the published journal article by providing the PDF file of the main text of the article. The output from the LLM chatbot provides lab-independent instructions for the experiment, which includes no information about specific instrumental implementation.
    \item Input the lab-independent instructions to the HAL system, with an extra leading prompt formatted as a query to generate a new experimental plan. The request will be directed to the HAL \textit{answer} component as discussed in the previous section. In the leading prompt, we can set preferences for implementation details such as data format, as well as refer to existing knowledge documents, such as ``IQ Scatter Experiment Plan'' and ``Coding Guide for QICK Board Experiment'', which are retrieved by the \textit{search} component and attached to the answering request. With a high-thinking configuration \cite{google2026gemini}, the \textit{answer} component can analyze the lab-independent instructions and incorporate implementation details based on the indicated documents. Emerging from existing knowledge, a lab-specific instruction (``QND Experiment Plan'') is generated, which is a practical guideline for the \textit{planner} to conduct the experiment from the journal article using our laboratory implementation and our specifications for instrumental control, variable format, data saving, etc.
    \item Add the lab-specific instructions to the knowledge base with a computed embedding vector for semantic search. From this point, the \textit{search} component can find this document in the knowledge base and provide it to other components when appropriate.
\end{enumerate}

Once the knowledge document has been prepared, HAL cycles are initiated with user input requesting a QND characterization with the provided QICK board address and calibrated qubit variables. In the first cycle, the \textit{planner} is supported by the \textit{search} component with the relevant documents, including the newly generated ``QND Experiment Plan'', which guides the \textit{planner} to prompt the \textit{developer} for the implementation of data acquisition using the experimental method described in the journal article \cite{hazra2025benchmarking}. The generated code is executed, the experiment is conducted using a QICK board, and the correlation data (Fig.~\ref{fig:qnd}(b), blue) is saved to the storage server. In the second cycle, the \textit{planner} instructs the \textit{developer} to load and fit the data against the model (Eq.~\ref{eq:qnd}), autonomously extracted from the journal article \cite{hazra2025benchmarking} and recorded in the plan document. A fit (Fig.~\ref{fig:qnd}(b), red) is performed by the generated code, and the leakage rate $L = 0.124\pm0.017$ is extracted. The details of the qubit performance are beyond the scope of this work. The observed data serve to confirm the successful implementation of the methods described in the journal article.

The relevant prompts, documents, data, a full transcript, and a screen record of this example can be found in the supplementary material \cite{Supplement}. The workflow including knowledge preparation uses on the order of $10^5$ input and $10^4$ output tokens (including thinking tokens). With prepared prompts, the process takes about three minutes, including the pulse sequence execution time.

The iterative RAG procedure implemented in the HAL \textit{search} component plays a critical role in this example, making it possible for HAL to generate and memorize dynamic knowledge. Relevant existing knowledge documents are provided to the \textit{answer} component, and the newly generated document can be instantaneously added to the knowledge base, for future use by HAL.

We emphasize the flexibility and reusability demonstrated by this example. Along with the convenience of reproducing a published experiment, leading prompts in the knowledge preparation process can be customized, inserting requested experimental details. Every artifact (lab-independent instructions, lab-specific instructions, prompts, codes) generated in the workflow can be modified and stored for customization and future use.

\begin{figure}
\includegraphics[width=\columnwidth]{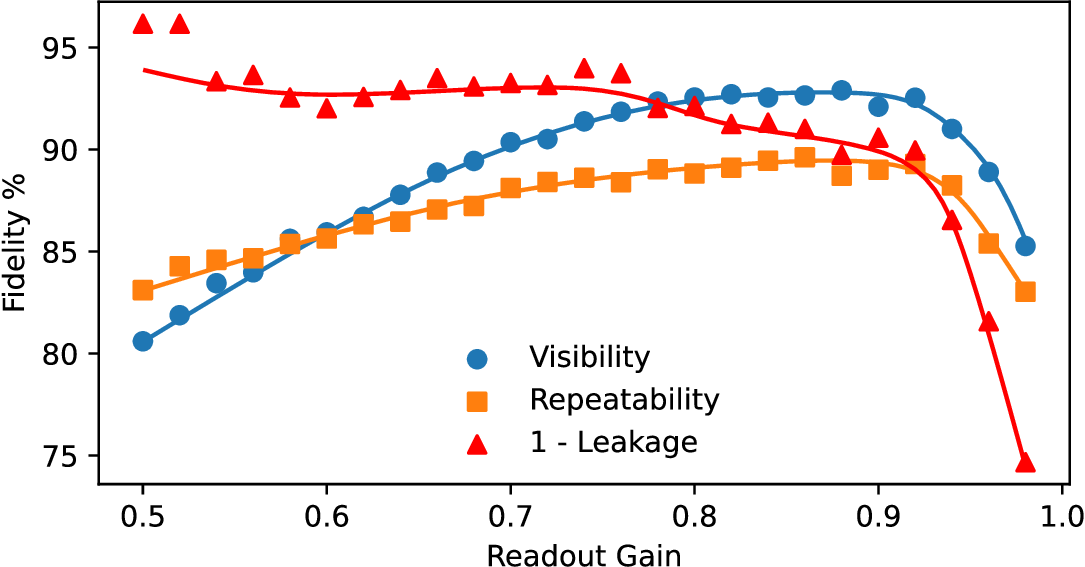}
\caption{\label{fig:fidelity} Readout fidelity metrics \cite{sunada2022fast,swiadek2024enhancing,hazra2025benchmarking} (visibility, repeatability, and complement of leakage ($1-L$)) as a function of readout power, the latter in units of the QICK board channel gain. The solid lines are polylines for visual guidance. The plot shows improvement of visibility and repeatability with readout power until near the maximum shown here, but general degradation due to increased leakage as power is increased.}
\end{figure}

We combine and modify the QND characterization code generated by HAL in this example with codes from other experiments (also generated by HAL), and use them to measure the readout fidelity metrics including visibility, repeatability, and complement of leakage ($1-L$) as a function of readout power. These results are displayed in Fig.~\ref{fig:fidelity}, which shows the measurement degradation due to increased leakage at high powers.

\section{Conclusion and Outlook}

In summary, we have built an innovative AI system driven by an LLM that provides expert assistance for superconducting qubit experiments. We describe two autonomous experiments that demonstrate the flexibility and convenience possible for superconducting resonator characterization and direct reproduction of an innovative experiment from a published journal article. The HAL system and its supporting open-source packages (QuICK, Grapher, etc.) show the promising ability to assist human researchers in superconducting circuit experiments.

We expect the advancement of LLMs to expand the potential of this AI system. We use a commercial LLM and rely on RAG for simplicity and transferability, avoiding the technical burden of local model deployment. This approach facilitates easy migration to newer reasoning models and allows for dynamic knowledge updates without retraining. For more complicated applications, better reasoning models (Gemini 3.1 Pro \cite{google2026gemini}, etc.) can be applied by reconfiguring HAL. Future large-scale implementations may, however, require fine-tuned models for specialized roles such as the \textit{planner} and \textit{developer}.

We recognize that a significant limitation of the HAL system is in its strong dependence on a knowledge base. This heuristic nature, while providing better human control, prevents the system from making voluntary actions to surpass human experience. The heuristic knowledge base also entails effort to collect laboratory-specific knowledge into formatted documents. However, we believe that this issue can be resolved by further development of automatic knowledge generation and synthesis. Future LLM agents may autonomously optimize the internal and external knowledge bases. We are optimistic that by combining literature review, web search, and review of the existing internal laboratory code base, an AI system can generate a better-organized knowledge base than human-provided documents. It may also identify and evaluate missing knowledge and spontaneously explore using the experimental setup with little human intervention. The HAL system could then serve as a control package for a higher-level AI system, becoming a cornerstone of a ``living'' lab.

We believe that this work can serve as the basis for future development of high-level AI systems to perform autonomous quantum experiments. Natural language interfaces allow for better monitoring and easier coordination across laboratory environments. Beyond pulse-level control, AI is already demonstrating utility in quantum algorithm design, circuit simulation, and fabrication \cite{alexeev2025artificial,mayor2025robotic}, potentially streamlining the transition from circuit design to measurement results, accelerating the research cycle in quantum science. We hope the design of the HAL system can also inspire more applications of LLMs on other innovative frontiers.

\begin{acknowledgments}
This work is supported by the Army Research Office and Laboratory for Physical Sciences (ARO grant W911NF2310077), the Air Force Office of Scientific Research (AFOSR grant FA9550-20-1-0270 and MURI grant FA9550-23-1-0338), DARPA DSO (grant HR0011-24-9-0364), and in part by UChicago's MRSEC (NSF award DMR-2011854), by the NSF QLCI for HQAN (NSF award 2016136), by the Simons Foundation (award 5099) and a 2024 Department of Defense Vannevar Bush Faculty Fellowship (ONR N000142512032). Results are in part based on work supported by the U.S. Department of Energy Office of Science National Quantum Information Science Research Centers as part of the Q-NEXT center. The authors declare no competing financial interests. Correspondence and requests for materials should be addressed to A. N. Cleland (anc@uchicago.edu).
\end{acknowledgments}

% \appendix
% \section{Appendixes}

\bibliography{main}

@PREAMBLE{
 "\providecommand{\noopsort}[1]{}" 
 # "\providecommand{\singleletter}[1]{#1}%" 
}

@article{google2025quantum,
  author={{Google Quantum AI and Collaborators}},
  title={Quantum error correction below the surface code threshold},
  journal={Nature},
  volume={638},
  number={8052},
  pages={920--926},
  year={2025},
  publisher={Nature Publishing Group UK London}
}

@misc{ibm2025qdc,
  author={{IBM Quantum}},
  title={Scaling for quantum advantage and beyond},
  howpublished={\url{https://www.ibm.com/quantum/blog/qdc-2025}},
  year={2025},
  month={November},
  note={Accessed: 2026-01-27}
}

@article{qiao2023splitting,
  title={Splitting phonons: Building a platform for linear mechanical quantum computing},
  author={Qiao, Hong and Dumur, {\'E}tienne and Andersson, Gustav and Yan, Haoxiong and Chou, M-H and Grebel, Joel and Conner, Christopher R and Joshi, Yash J and Miller, Jacob M and Povey, Rhys G and others},
  journal={Science},
  volume={380},
  number={6649},
  pages={1030--1033},
  year={2023},
  publisher={American Association for the Advancement of Science}
}

@article{qiao2025acoustic,
  title={Acoustic phonon phase gates with number-resolving phonon detection},
  author={Qiao, Hong and Wang, Zhaoyou and Andersson, Gustav and Anferov, Alexander and Conner, Christopher R and Joshi, Yash J and Li, Shiheng and Miller, Jacob M and Wu, Xuntao and Yan, Haoxiong and others},
  journal={Nature Physics},
  pages={1--5},
  year={2025},
  publisher={Nature Publishing Group UK London}
}

@article{wu2024modular,
  title={Modular quantum processor with an all-to-all reconfigurable router},
  author={Wu, Xuntao and Yan, Haoxiong and Andersson, Gustav and Anferov, Alexander and Chou, Ming-Han and Conner, Christopher R and Grebel, Joel and Joshi, Yash J and Li, Shiheng and Miller, Jacob M and others},
  journal={Physical Review X},
  volume={14},
  number={4},
  pages={041030},
  year={2024},
  publisher={APS}
}

@article{wu2025mitigating,
  title={Mitigating cosmic-ray-like correlated events with a modular quantum processor},
  author={Wu, Xuntao and Joshi, Yash J and Yan, Haoxiong and Andersson, Gustav and Anferov, Alexander and Conner, Christopher R and Karimi, Bayan and King, Amber M and Li, Shiheng and Malc, Howard L and others},
  journal={Physical Review Applied},
  volume={24},
  number={4},
  pages={044022},
  year={2025},
  publisher={APS}
}

@article{zhu2011coherent,
  title={Coherent coupling of a superconducting flux qubit to an electron spin ensemble in diamond},
  author={Zhu, Xiaobo and Saito, Shiro and Kemp, Alexander and Kakuyanagi, Kosuke and Karimoto, Shin-ichi and Nakano, Hayato and Munro, William J and Tokura, Yasuhiro and Everitt, Mark S and Nemoto, Kae and others},
  journal={Nature},
  volume={478},
  number={7368},
  pages={221--224},
  year={2011},
  publisher={Nature Publishing Group UK London}
}

@article{scarlino2019coherent,
  title={Coherent microwave-photon-mediated coupling between a semiconductor and a superconducting qubit},
  author={Scarlino, Pasquale and Van Woerkom, David J and Mendes, Udson C and Koski, Jonne V and Landig, Andreas J and Andersen, Christian Kraglund and Gasparinetti, Simone and Reichl, Christian and Wegscheider, Werner and Ensslin, Klaus and others},
  journal={Nature communications},
  volume={10},
  number={1},
  pages={3011},
  year={2019},
  publisher={Nature Publishing Group UK London}
}

@article{mirhosseini2020superconducting,
  title={Superconducting qubit to optical photon transduction},
  author={Mirhosseini, Mohammad and Sipahigil, Alp and Kalaee, Mahmoud and Painter, Oskar},
  journal={Nature},
  volume={588},
  number={7839},
  pages={599--603},
  year={2020},
  publisher={Nature Publishing Group UK London}
}

@article{clerk2020hybrid,
  title={Hybrid quantum systems with circuit quantum electrodynamics},
  author={Clerk, AA and Lehnert, KW and Bertet, P and Petta, JR and Nakamura, Y},
  journal={Nature Physics},
  volume={16},
  number={3},
  pages={257--267},
  year={2020},
  publisher={Nature Publishing Group UK London}
}

@article{hou2025model,
  title={Model context protocol (mcp): Landscape, security threats, and future research directions},
  author={Hou, Xinyi and Zhao, Yanjie and Wang, Shenao and Wang, Haoyu},
  journal={arXiv preprint arXiv:2503.23278},
  year={2025}
}

@article{minaee2024large,
  title={Large language models: A survey},
  author={Minaee, Shervin and Mikolov, Tomas and Nikzad, Narjes and Chenaghlu, Meysam and Socher, Richard and Amatriain, Xavier and Gao, Jianfeng},
  journal={arXiv preprint arXiv:2402.06196},
  year={2024}
}

@article{tom2024self,
  title={Self-driving laboratories for chemistry and materials science},
  author={Tom, Gary and Schmid, Stefan P and Baird, Sterling G and Cao, Yang and Darvish, Kourosh and Hao, Han and Lo, Stanley and Pablo-Garc{\'\i}a, Sergio and Rajaonson, Ella M and Skreta, Marta and others},
  journal={Chemical Reviews},
  volume={124},
  number={16},
  pages={9633--9732},
  year={2024},
  publisher={ACS Publications}
}

@article{swanson2025virtual,
  title={The Virtual Lab of AI agents designs new SARS-CoV-2 nanobodies},
  author={Swanson, Kyle and Wu, Wesley and Bulaong, Nash L and Pak, John E and Zou, James},
  journal={Nature},
  volume={646},
  number={8085},
  pages={716--723},
  year={2025},
  publisher={Nature Publishing Group UK London}
}

@article{cao2025automating,
  title={Automating quantum computing laboratory experiments with an agent-based AI framework},
  author={Cao, Shuxiang and Zhang, Zijian and Alghadeer, Mohammed and Fasciati, Simone D and Piscitelli, Michele and Bakr, Mustafa and Leek, Peter and Aspuru-Guzik, Al{\'a}n},
  journal={Patterns},
  volume={6},
  number={10},
  year={2025},
  publisher={Elsevier}
}

@article{krantz2019quantum,
  title={A quantum engineer's guide to superconducting qubits},
  author={Krantz, Philip and Kjaergaard, Morten and Yan, Fei and Orlando, Terry P and Gustavsson, Simon and Oliver, William D},
  journal={Applied physics reviews},
  volume={6},
  number={2},
  year={2019},
  publisher={AIP Publishing}
}

@article{blais2021circuit,
  title={Circuit quantum electrodynamics},
  author={Blais, Alexandre and Grimsmo, Arne L and Girvin, Steven M and Wallraff, Andreas},
  journal={Reviews of Modern Physics},
  volume={93},
  number={2},
  pages={025005},
  year={2021},
  publisher={APS}
}

@article{krinner2019engineering,
  title={Engineering cryogenic setups for 100-qubit scale superconducting circuit systems},
  author={Krinner, Sebastian and Storz, Simon and Kurpiers, Philipp and Magnard, Paul and Heinsoo, Johannes and Keller, Raphael and Luetolf, Janis and Eichler, Christopher and Wallraff, Andreas},
  journal={EPJ Quantum Technology},
  volume={6},
  number={1},
  pages={2},
  year={2019},
  publisher={Springer Berlin Heidelberg}
}

@article{hazra2025benchmarking,
  title={Benchmarking the readout of a superconducting qubit for repeated measurements},
  author={Hazra, S and Dai, W and Connolly, T and Kurilovich, PD and Wang, Z and Frunzio, L and Devoret, MH},
  journal={Physical Review Letters},
  volume={134},
  number={10},
  pages={100601},
  year={2025},
  publisher={APS}
}

@article{stefanazzi2022qick,
  title={The QICK (Quantum Instrumentation Control Kit): Readout and control for qubits and detectors},
  author={Stefanazzi, Leandro and Treptow, Kenneth and Wilcer, Neal and Stoughton, Chris and Bradford, Collin and Uemura, Sho and Zorzetti, Silvia and Montella, Salvatore and Cancelo, Gustavo and Sussman, Sara and others},
  journal={Review of Scientific Instruments},
  volume={93},
  number={4},
  year={2022},
  publisher={AIP Publishing}
}

@misc{google2026gemini,
  title={Gemini 3 Developer Guide},
  howpublished={\url{https://ai.google.dev/gemini-api/docs/gemini-3}},
  year={2026},
  note={Accessed: 2026-01-26}
}

@misc{google2026gemini_web,
  author = {{Google}},
  title = {Gemini},
  year = {2026},
  url = {https://gemini.google.com/},
  note = {Accessed: February 8, 2026}
}

@article{gao2023retrieval,
  title={Retrieval-augmented generation for large language models: A survey},
  author={Gao, Yunfan and Xiong, Yun and Gao, Xinyu and Jia, Kangxiang and Pan, Jinliu and Bi, Yuxi and Dai, Yixin and Sun, Jiawei and Wang, Haofen and Wang, Haofen},
  journal={arXiv preprint arXiv:2312.10997},
  volume={2},
  number={1},
  year={2023}
}

@article{laban2025llms,
  title={Llms get lost in multi-turn conversation},
  author={Laban, Philippe and Hayashi, Hiroaki and Zhou, Yingbo and Neville, Jennifer},
  journal={arXiv preprint arXiv:2505.06120},
  year={2025}
}

@article{grecco2023pyvisa,
  author = {Grecco, Hernán E. and Dartiailh, Matthieu C. and Thalhammer-Thurner, Gregor and Bronger, Torsten and Bauer, Florian},
  doi = {10.21105/joss.05304},
  journal = {Journal of Open Source Software},
  month = {4},
  number = {84},
  pages = {5304--5308},
  title = {{PyVISA: the Python instrumentation package}},
  volume = {8},
  year = {2023}
}

@misc{github_grapher,
  title={{Grapher} {G}ithub repository},
  year={2026},
  publisher={GitHub},
  journal={GitHub repository},
  howpublished={\url{https://github.com/clelandlab/grapher}}
}

@misc{github_quick,
  title={{QuICK} {G}ithub repository},
  year={2026},
  publisher={GitHub},
  journal={GitHub repository},
  howpublished={\url{https://github.com/clelandlab/quick}}
}

@misc{github_hal,
  title={{HAL} {G}ithub repository},
  year={2026},
  publisher={GitHub},
  journal={GitHub repository},
  howpublished={\url{https://github.com/clelandlab/HAL}}
}

@article{megrant2012planar,
  title={Planar superconducting resonators with internal quality factors above one million},
  author={Megrant, Anthony and Neill, Charles and Barends, Rami and Chiaro, Ben and Chen, Yu and Feigl, Ludwig and Kelly, Julian and Lucero, Erik and Mariantoni, Matteo and O’Malley, Peter JJ and others},
  journal={Applied Physics Letters},
  volume={100},
  number={11},
  year={2012},
  publisher={AIP Publishing}
}

@article{sunada2022fast,
  title={Fast readout and reset of a superconducting qubit coupled to a resonator with an intrinsic Purcell filter},
  author={Sunada, Yoshiki and Kono, Shingo and Ilves, Jesper and Tamate, Shuhei and Sugiyama, Takanori and Tabuchi, Yutaka and Nakamura, Yasunobu},
  journal={Physical Review Applied},
  volume={17},
  number={4},
  pages={044016},
  year={2022},
  publisher={APS}
}

@article{swiadek2024enhancing,
  title={Enhancing dispersive readout of superconducting qubits through dynamic control of the dispersive shift: Experiment and theory},
  author={Swiadek, Fran{\c{c}}ois and Shillito, Ross and Magnard, Paul and Remm, Ants and Hellings, Christoph and Lacroix, Nathan and Ficheux, Quentin and Zanuz, Dante Colao and Norris, Graham J and Blais, Alexandre and others},
  journal={PRX Quantum},
  volume={5},
  number={4},
  pages={040326},
  year={2024},
  publisher={APS}
}

@article{mosqueira2023human,
  title={Human-in-the-loop machine learning: a state of the art},
  author={Mosqueira-Rey, Eduardo and Hern{\'a}ndez-Pereira, Elena and Alonso-R{\'\i}os, David and Bobes-Bascar{\'a}n, Jos{\'e} and Fern{\'a}ndez-Leal, {\'A}ngel},
  journal={Artificial Intelligence Review},
  volume={56},
  number={4},
  pages={3005--3054},
  year={2023},
  publisher={Springer}
}

@article{alexeev2025artificial,
  title={Artificial intelligence for quantum computing},
  author={Alexeev, Yuri and Farag, Marwa H and Patti, Taylor L and Wolf, Mark E and Ares, Natalia and Aspuru-Guzik, Al{\'a}n and Benjamin, Simon C and Cai, Zhenyu and Cao, Shuxiang and Chamberland, Christopher and others},
  journal={Nature Communications},
  volume={16},
  number={1},
  pages={10829},
  year={2025},
  publisher={Nature Publishing Group UK London}
}

@article{mayor2025robotic,
  title={Robotic chip-scale nanofabrication for superior consistency},
  author={Mayor, Felix M and Guan, Wenyan and Szakiel, Erik and Safavi-Naeini, Amir H and Gyger, Samuel},
  journal={arXiv preprint arXiv:2511.19432},
  year={2025}
}

@Article{Supplement,
  author = {},
  title = {Supplemental Material},
  year = 2026,
  journal   = {}
}

\end{document}